\documentclass[%
 reprint,
aps,
nofootinbib]{revtex4-1}
\usepackage{scrextend}
\usepackage{graphicx}
\usepackage{dcolumn}
\usepackage{bm}
\usepackage{float}
\usepackage{hyperref}
\usepackage[mathlines]{lineno}
\usepackage{footnote}
\usepackage{amsmath}
\usepackage{xfrac}
\usepackage{braket}
\usepackage{cancel}
\usepackage{xcolor}
\usepackage{amssymb}
\usepackage[mathscr]{euscript}
\usepackage{lipsum}
\usepackage{mathtools}
\usepackage{cuted}
\usepackage[export]{adjustbox}

\usepackage{mathtools}

\newcommand{\arXivold}[2]{\href{http://arxiv.org/pdf/#1}{{\tt #2/#1}}}

\usepackage{blindtext} 

\begin{document}

\preprint{APS/123-QED}

\title{A Proposed Renormalization Scheme for Non-local QFTs and Application to the Hierarchy Problem}
\author{Fayez Abu-Ajamieh}
 \email{fayezajamieh@iisc.ac.in}
\affiliation{%
Centre for High Energy Physics, Indian Institute of Science, Bangalore 560012, India
}
\author{Sudhir K. Vempati}
\email{vempati@iisc.ac.in}
\affiliation{Centre for High Energy Physics, Indian Institute of Science, Bangalore 560012, India}


\begin{abstract}
We propose a renormalization scheme for non-local Quantum Field Theories (QFTs) with infinite derivatives inspired by string theory. Our Non-locality Renormalization Scheme (NRS) is inspired by Dimensional Regularization (DR) in local QFTs and is shown to significantly improve the UV behavior of non-local QFTs. We illustrate the scheme using simple examples from the $\phi^{3}$ and $\phi^{4}$ theories, then we evaluate the viability of NRS-enhanced non-local QFTs to solve the hierarchy problem using a simplified toy model. We find that non-locality protects the mass of a light scalar from receiving large corrections from any UV sector to which it couples, as long as the non-locality scale $\Lambda$ is \textit{sufficiently} smaller than the scale of the UV sector. We also find that NRS eliminates any large threshold corrections from the IR sector.

\end{abstract}
\pacs{Valid PACS appear here}
\maketitle


\section{Introduction}\label{sec:Introduction}
It is commonly known that (local) QFTs are plagued with UV divergences, which emanate from the point-like nature of local interactions. Such locality renders QFTs sensitive to very short distances/high frequencies, leading to uncontrolled growth in scattering amplitudes. The standard prescription for dealing with such divergences is to utilize a renormalization scheme, such as DR, whereby one subtracts these divergences essentially by hand. On the other hand, strings are naturally free of UV divergences owing to their inherently non-local nature. The finite size of strings provides a natural cutoff for very short distances/high frequencies which regularizes UV divergences without the need for any renormalization scheme. 

Inspired by the good UV behavior of scattering amplitudes of strings \cite{Witten:1985cc, Seiberg:1999vs, Kostelecky:1989nt, Kostelecky:1988ta, Freund:1987kt, Freund:1987ck, Brekke:1987ptq, Frampton:1988kr, Tseytlin:1995uq, Biswas:2004qu, Siegel:2003vt}, non-local QFTs with infinite derivatives \cite{Biswas:2014yia} attempt at mimicking string theories by introducing non-locality to the propagators of local QFTs via the exponent of an entire function suppressed by the scale of non-locality. For example, a non-local (real) scalar QFT takes the form  
\begin{equation}\label{eq:phi4Lag}
\mathscr{L}_{\phi} = -\frac{1}{2}\phi e^{\frac{\Box +m^{2}}{\Lambda^{2}}}(\Box +m^{2})\phi - \frac{\lambda}{4!}\phi^{4},
\end{equation}
where $\Box = \partial_{\mu}\partial^{\mu}$ and $\Lambda$ is the scale of non-locality. Notice that that when $\Lambda \rightarrow \infty$, we retrieve the local QFT. The non-local form factor in eq. (\ref{eq:phi4Lag}) has several attractive features: It  does not introduce any new poles to the local propagator, which means that the particle spectrum in the non-local QFTs is identical to the local case. It also makes the connection with the Lee-Wick (LW) theory \cite{Lee:1969fy, Lee:1970iw, Grinstein:2007mp} transparent, as the LW theory can simply be obtained by expanding the form factor and keeping the leading derivative operator. More importantly, the derivative in the exponent improves the UV behavior of the theory, as when calculating loop diagrams, the form factor in momentum space becomes $\sim e^{-k^{2}/\Lambda^{2}}$, which suppresses all large momenta above the non-locality scale, thereby providing a physical cutoff for the theory. Therefore, such non-local QFTs are expected to be super-renormalizable. This feature was used to show that the $\beta$-functions of Abelian \cite{Ghoshal:2017egr} and non-Abelian \cite{ Ghoshal:2020lfd} non-local QFTs exhibit a conformal behavior and flow towards fixed points, thereby providing a possible solution to the Higgs potential instability problem \cite{Cabibbo:1979ay,Hung:1979dn, Lindner:1985uk,Sher:1988mj,Schrempp:1996fb,Altarelli:1994rb, Degrassi:2012ry, Buttazzo:2013uya}. In addition, it was speculated that such a formulation has the potential of solving the hierarchy problem if $\Lambda \sim O(\text{TeV})$ \cite{Krasnikov:1987yj, Moffat:1988zt, Moffat:1990jj}.

On the other hand, eq. (\ref{eq:phi4Lag}) is not free of problems. First, eq. (\ref{eq:phi4Lag}) is an ansatz that is not derived from any first principles. Thus, it should be treated as an Effective Field Theory (EFT) of yet a more fundamental theory in the UV\footnote{Such a form factor could arise from the star product in noncommutative geometries, see for instance \cite{Filk:1996dm, Varilly:1998gq, Chaichian:1998kp, Chaichian:1999wy, Sheikh-Jabbari:1999olp, Martin:1999aq, Krajewski:1999ja, Cho:1999sg, Hawkins:1999gn, Bigatti:1999iz, Ishibashi:1999hs, Chepelev:1999tt, Benaoum:1999ca, Minwalla:1999px}.}. In addition, there is no reason why the non-local form factor should take the form given in eq. (\ref{eq:phi4Lag}), and other form factor could in principle lead to an altered behavior (see \cite{BasiBeneito:2022wux} for a review). Perhaps most embarrassingly, any non-local QFT will lead to acausality (see for instance \cite{Coleman, Alvarez:2009af}), albeit at the microscopic level.\footnote{It is possible for causality in such theories to be an emergent property at the macroscopic level, see for instance \cite{Grinstein:2008bg}}. In spite of all these reservations, the formulation in eq. (\ref{eq:phi4Lag}) remains attractive as an EFT that improves the UV behavior of scattering amplitudes, in addition to circumventing the ill-defined notion of point interactions. 

Expressing non-locality as the exponent of an entire function as in eq (\ref{eq:phi4Lag}), was supposed to render the theory super-renormalizable without the need to resort to any renormalization scheme, however, we shall show in this letter that for certain types of integrals, amplitudes would receive large threshold corrections from the non-locality scale, which become even larger with increasing the non-locality scale and eventually divergent when $\Lambda \rightarrow \infty$. Such a behavior is counter-intuitive, as one would expect the effects of any UV physics to yield subleading corrections to the IR sector in any well-behaved EFT, and eventually decouple completely when $\Lambda \rightarrow \infty$. 

Given that the - so far - null results in the LHC seem to suggest that the scale of non-locality might be larger than what was once expected (assuming it exists), it seems that such large corrections are unavoidable. We are thus faced with one of two options: Either we give up on non-locality, or we enhance non-local QFTs with a proper renormalization scheme that cures their ills. We show that augmenting non-local QFTs with a proper renormalization scheme would eliminate any large corrections to amplitudes that are sensitive to the scale of non-locality, and we argue that such renormalization should be part of the definition of a particle's wavefunction. We formulate our proposed renormalization scheme by drawing inspiration from DR and show that it strengthens the non-local QFTs framework. 

The necessity of renormalizing non-local QFTs was recognized in \cite{Capolupo:2022awe}, where non-local QED was proposed as a solution to the $g-2$ anomaly. In addition, renormalization of non-local QFTs was touched on in \cite{Biswas:2014yia}, where they renormalized the 1-loop $\phi\phi \rightarrow \phi\phi$ scattering amplitude by expanding the form factor and subtracting the divergent modes in the expansion. However, their approach is somewhat ad-hoc, and was only applied to the aforementioned scattering. In our approach, we formulate a more rigorous renormalization scheme that can be applied to any theory. In addition, we provide more motivation for this renormalization and show how it connects to local QFTs. Our results are reminiscent of the findings in \cite{Minwalla:1999px}, where a UV cutoff naturally arises from the noncommutativity of spacetime, thereby regularizing certain types of diagrams, however, as explained therein, that only occurs in nonplanar diagrams, whereas divergent planar diagrams still need a regularization scheme.

This paper is organized as follows: In Section \ref{sec:Motivation} we present the motivation for formulating a renormalization scheme for non-local QFTs. In Section \ref{sec:Scheme} we define the renormalization scheme and illustrate it using simple examples from the $\phi^{3}$ and $\phi^{4}$ theories, and as a check, we show that the $\beta$-function of the $\phi^{4}$ theory remains unchanged. In Section \ref{sec:HierarchyProblem} we apply the non-local renormalization scheme to the hierarchy problem and show that it enhances the viability of non-local QFTs as a potential solution, and finally, we present our conclusions and discuss the outlook in Section \ref{Sec:conclusions}.

\section{Motivation}\label{sec:Motivation}
\subsection{Motivation from the $\phi^{4}$ 2-point Function}\label{sec:motivation1}
We begin motivating the renormalization scheme by considering the 1-loop correction to the 2-point function in the $\phi^{4}$ in eq. (\ref{eq:phi4Lag}). First, we consider the local case by taking $\Lambda \rightarrow \infty$ in the Lagrangian, which gives us the familiar local case. In $d=4-2\epsilon$, the 1-loop correction is given by
\begin{equation}\label{eq:2ptLocal1}
\Pi_{2} = \frac{\lambda m^{2}}{32\pi^{2}} \Big[ \frac{1}{\epsilon}-\gamma_{E} +1+ \log\Big( \frac{4\pi \mu^{2}}{m^{2}}\Big) + O(\epsilon) \Big],
\end{equation}
and we see that the result is divergent. However, the result is rendered finite by subtracting the divergence with an appropriate counterterm. For instance, in the $\overline{\text{MS}}$ scheme, we have
\begin{equation}\label{eq:2ptLocal2}
\Pi_{2} = \frac{\lambda m^{2}}{32\pi^{2}} \Big[1 - \log \Big( \frac{m^{2}}{\mu^{2}}\Big)\Big],
\end{equation}
where $\mu$ is the renormalization scale, usually taken to be the scale of relevant process. The non-local theory, on the other hand, is expected to be super-renormalizable, removing the divergence directly. However, the 2-point function in the non-local theory is found to be
\begin{equation}\label{eq:2ptnon-local1}
\Pi_{2}^{\text{NL}} = -\frac{\lambda}{32\pi^{2}}e^{-\frac{m^{2}}{\Lambda^{2}}} \Big[ \Lambda^{2} + m^{2} e^{\frac{m^{2}}{\Lambda^{2}}} \text{Ei}\Big(\frac{-m^{2}}{\Lambda^{2}}\Big) \Big],
\end{equation}
where the exponential integral function $\text{Ei}(x)$ is defined as
\begin{equation}\label{eq:Ei}
\text{Ei}(x) = - \int_{-x}^{\infty}dt \frac{e^{-t}}{t},
\end{equation}
with the Cauchy principal value taken. For a small argument $z$, $\text{Ei}(z)$ behaves like a logarithmic function. 

Inspecting eq.(\ref{eq:2ptnon-local1}), we can see that the non-local 2-point function receives large corrections when the scale of non-locality is large, and diverges when $\Lambda \rightarrow \infty$ both quadratically through the $\Lambda^{2}$ term, and logarithmically through the $\text{Ei}(\frac{-m^{2}}{\Lambda^{2}})$ function. Therefore, unless $\Lambda$ is not much larger than $m$, the corrections will be unacceptably large. For instance, if we treat $\phi$ as a Higgs-like particle with $m \sim O(\text{EW})$, then unless $\Lambda \sim$ a few TeV, the correction to the mass of $\phi$ will be too large, in spite of the fact that in the renormalized local case, these corrections are under control. Thus, if $\Lambda$ is large, then it will make things worse so far as calculating mass corrections is concerned, instead of ameliorating their divergences.

To better illustrate this issue, notice that although introducing non-locality to the Lagrangian does indeed remove the divergences and make amplitudes finite, nevertheless, the divergences in the unrenormalized local theory translate into "pseudo-divergences" in the non-local theory. Specifically, the $\epsilon^{-1}$ factor (which corresponds to a quadratic divergence had we used a UV cutoff instead of DR), translates into $\Lambda^{2}$, whereas the logarithmic divergence translates into the exponential integral function $\text{Ei}(\frac{-m^{2}}{\Lambda^{2}})$. This seems to suggest that non-locality alone is insufficient to remove UV divergences fully, especially if we allow scale of non-locality to be high.

\subsection{Motivation from Oblique Parameters}\label{sec:motivations2}
Next, we consider the contributions of non-locality to the STU parameters. To keep things simple, we consider the corrections due to the Higgs in a non-local QFT.\footnote{We should note that the non-local versions of the EW sector and Higgs mechanism have not been fully developed yet and still face issues, such as the putative existence of ghost degrees of freedom (see for instance \cite{Hashi:2018kag} and the proposed solution \cite{Buoninfante:2018mre}). Nevertheless, we can ignore these issues and proceed by placing the non-locality on the Higgs propagator and assuming local interactions with gauge fields.} The Higgs contributes to the oblique parameters through modifying $\Pi_{WW}$ and $\Pi_{ZZ}$ as shown in Fig. \ref{fig1} below.

\begin{figure}[!h] 
\centering
\includegraphics[width=0.4\textwidth]{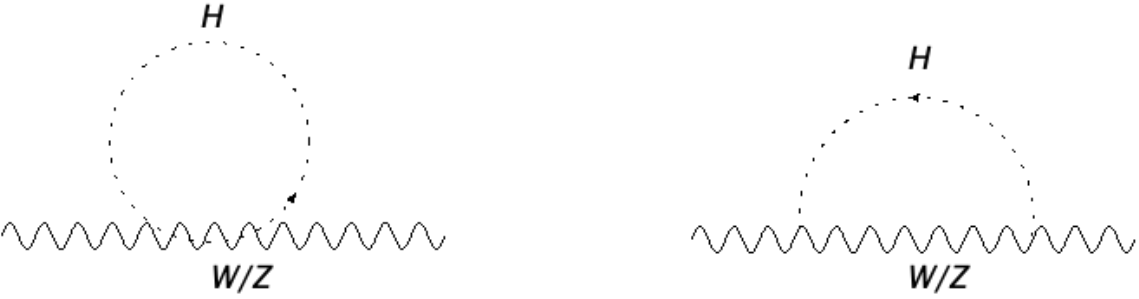}
\caption{Higgs corrections to $\Pi_{WW}$ and $\Pi_{ZZ}$ at 1-loop.}
\label{fig1}
\end{figure}

The left diagram can be evaluated exactly, yielding the following contribution
\begin{equation}\label{eq:PiVV1}
\Pi_{\text{VV}}^{(1)}(p^{2}) = \frac{g_{V}^{2}}{64\pi^{2}} e^{-\frac{m_{h}^{2}}{\Lambda^{2}}} \Big[ \Lambda^{2} + m_{h}^{2}e^{\frac{m_{h}^{2}}{\Lambda^{2}}} \text{Ei}\Big(\frac{-m_{h}^{2}}{\Lambda^{2}} \Big) \Big],
\end{equation}
where $g_{W}^{2} = g^{2}$, $g_{Z}^{2} = g^{2}+g'^{2}$ and $m_{h}$ is the Higgs mass $= 125$ GeV. The diagram on the right is more complicated, however, in the limit of a vanishing external momentum (valid for calculating the $T$ parameter for instance), it simplifies greatly and can be done exactly. Therefore, we focus on this case and limit ourselves to calculating the $T$ parameter, which is sufficient to illustrate our point. At vanishing external momentum, the second diagram yields
\begin{multline}\label{eq:PiVV2}
\Pi_{\text{VV}}^{(2)}(p^{2}=0) =  \frac{v^{2}g_{V}^{2}}{64\pi^{2}}\Big( \frac{1}{m_{V}^{2}-m_{h}^{2}}\Big) \\
 \times \Big[ m_{V}^{2}e^{\frac{m_{V}^{2}-m_{h}^{2}}{\Lambda^{2}}}\text{Ei}\Big(\frac{-2m_{V}^{2}}{\Lambda^{2}}\Big) - m_{h}^{2}e^{\frac{m_{h}^{2}-m_{V}^{2}}{\Lambda^{2}}}\text{Ei}\Big(\frac{-2m_{h}^{2}}{\Lambda^{2}}\Big) \Big],
\end{multline}
and the $T$ parameter is defined as
\begin{equation}\label{eq:Tparam}
T \equiv \frac{1}{\alpha}\Big( \frac{\Pi_{WW}(0)}{m_{W}^{2}} - \frac{\Pi_{ZZ}(0)}{m_{Z}^{2}} \Big).
\end{equation}

\begin{figure}[!t] 
\centering
\includegraphics[width=0.3\textwidth]{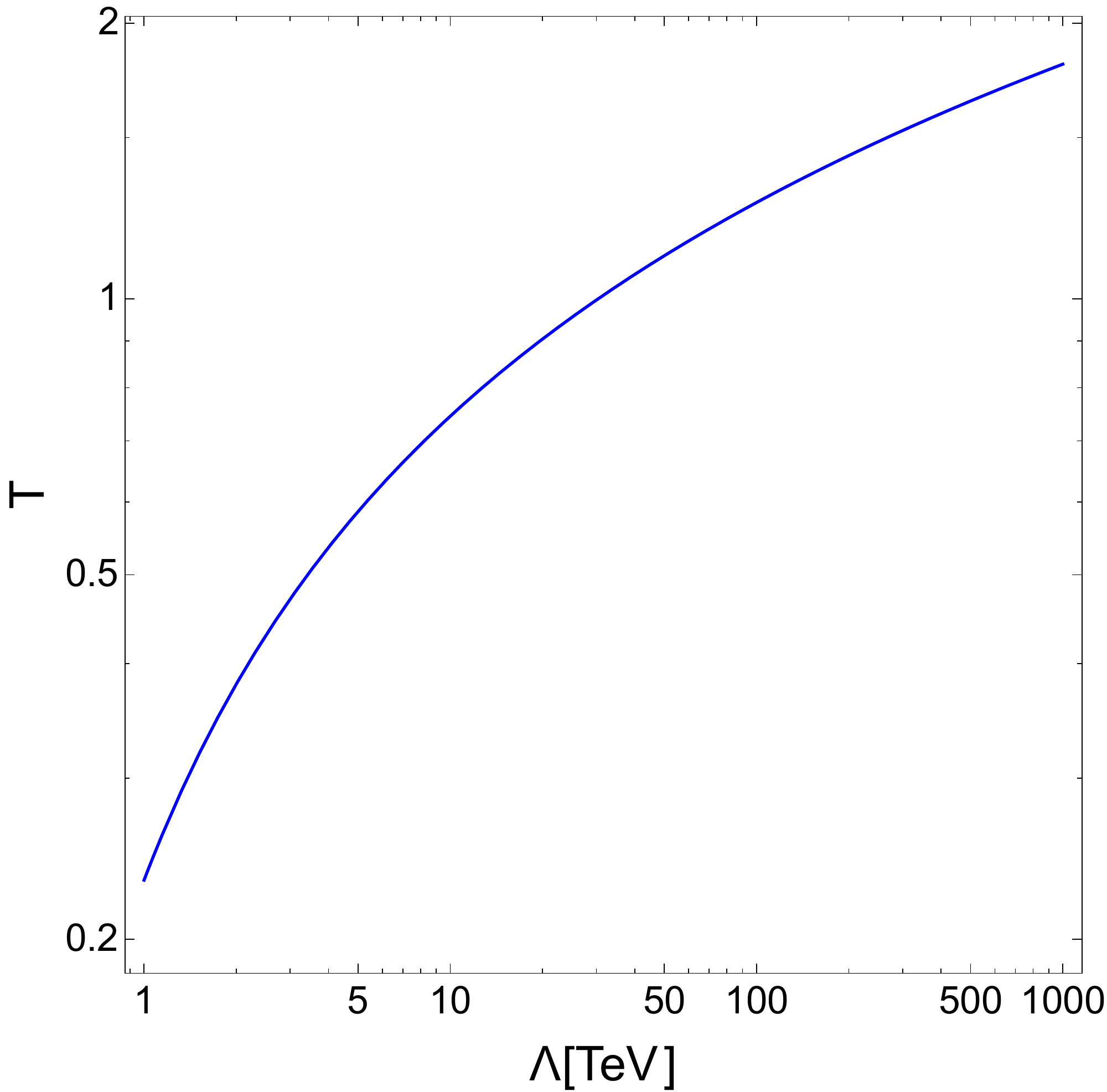}
\caption{The Higgs contribution to the $T$ parameter in a non-local QFT as a function of the non-locality scale $\Lambda$.}
\label{fig2}
\end{figure}

Inspecting eq. (\ref{eq:PiVV1}), we can easily see that $\Pi_{\text{VV}}^{(1)}$ yields a vanishing contribution to the $T$ parameter. On the other hand, the contribution from $\Pi_{\text{VV}}^{(2)}$ is non-vanishing. We plot this contribution in Fig. \ref{fig2} against the non-locality scale. The plot shows that the correction to the $T$ parameter is unacceptably large. For instance, the LHC constrains $\Lambda \gtrsim 2.5 - 3$ TeV \cite{Biswas:2004qu}. This corresponds to a $T$ parameter of $\sim 0.43 -0.47$, which is already excluded! Furthermore, the $T$ parameter increases with the scale of non-locality (albeit logarithmically), which is both counter-intuitive and unphysical. The reason why this behavior is unacceptable is that for an EFT to be sound, the IR sector should be insensitive to the UV physics, i.e., any UV physics should, once integrated out, be represented by an irrelevant operator (or operators) yielding suppressed contributions that become \textit{smaller} the higher the UV scale, eventually decoupling completely when the UV scale is sent to infinity. However, what we are witnessing here is exactly the opposite behavior!

Inspecting eq. (\ref{eq:PiVV2}), it is easy to see that this behavior originates from the exponential integral function $\text{Ei}(x)$. Expanding $\text{Ei}(x)$ for small (real) arguments, we find
\begin{equation}\label{eq:EiExpansion}
\text{Ei}\Big(\frac{-m^{2}}{\Lambda^{2}}\Big) \simeq \gamma_{\text{E}} + \log{\Big(\frac{m^{2}}{\Lambda^{2}}\Big)} + O\Big(\frac{m^{2}}{\Lambda^{2}}\Big),
\end{equation}
which diverges logarithmically for $\Lambda \rightarrow \infty$.

The two previous examples have shown us that for certain types of momentum integrals, specifically ones where the local QFT are either quadratically or logarithmically divergent, although non-locality does indeed render such integrals finite, it nonetheless does not protect IR observables from large corrections from the non-local UV sector. Specifically, the quadratic UV divergence in the local case translates into a quadratic dependence on the scale of non-locality, while logarithmic divergences in the local QFT translate into a logarithm-like dependence of the scale of non-locality through the exponential integral function $\text{Ei(x)}$. Therefore, we are faced with one of two options: 1) either we abandon non-locality as a viable UV completion for local QFTs (at least in the form presented here through an infinite derivative form factor), or 2) we enhance non-local QFTs with an appropriate renormalization scheme to remove the non-local divergence-like terms. We propose a non-local renormalization scheme to eliminate the terms that diverge in the limit $\Lambda \rightarrow \infty$ in the next section.

\section{non-locality Renormalization Scheme (NRS)}\label{sec:Scheme}
As we saw in the previous section, although non-locality does indeed cut-off UV divergences, it nonetheless results in unacceptably large threshold corrections to IR observables. In addition, the quadratic (and potentially logarithmic) dependence on the non-locality scale makes non-local QFTs unsound as an EFT description. We thus propose to remedy this situation by formulating a renormalization scheme for non-local QFTs that is inspired by DR in local QFTs.

When applying DR to local QFTs, divergences (such as in eq. (\ref{eq:2ptLocal1})) are subtracted by counterterms. When only the divergence is subtracted (i.e. $\epsilon^{-1}$), the scheme is called Minimum Subtraction (MS), whereas when one subtracts $\epsilon^{-1} + \gamma_{\text{E}} +\log{(4\pi)}$, the scheme is dubbed Modified Minimum Subtraction or $\overline{\text{MS}}$. The physical interpretation of this procedure is that the quantities appearing in the Lagrangian, which are called bare quantities, contain an infinite piece that is unobservable and should be part of the definition of the normalization of the quantity, and thus should be subtracted leaving behind the observable part only. For example, in QED, the bare charge can never be observed, as at short distances around it the vacuum becomes polarized, thereby screening the bare charge and rendering only the screened (physical) charge as the observable part.

Inspired by DR, we formulate the NRS as follows:
\begin{enumerate}

\item Quantities in the non-local Lagrangian are considered bare quantities that are physically unobservable, with the non-local divergence-like contributions to be removed by appropriate counterterms,

\item Any quadratic dependence on the non-locality scale $\Lambda$ should be absent from all physical observables,

\item The logarithmic dependence on $\Lambda$ may or may not be subtracted, depending on the scheme. If the logarithmic dependence is to be subtracted, then we do so via the following prescription
\begin{equation}\label{eq:logSub}
	\text{Ei}\Big(\frac{-m^{2}}{\Lambda^{2}}\Big) \rightarrow \text{Ei}\Big(\frac{-m^{2}}{\Lambda^{2}}\Big) - \text{Ei}\Big(\frac{-\mu^{2}}{\Lambda^{2}}\Big),
\end{equation}
with $\mu$ to be identified with the renormalization scale, just like in DR.

\item In the limit $\Lambda \rightarrow \infty$, the renormalized non-local quantity should agree with the corresponding dimensionally regularized local quantity up to possibly scheme-dependent constant terms.

\end{enumerate}
	
We call applying condition (1) alone the Minimum Non-locality Subtraction (MNS) scheme, and we call applying both conditions (1) and (3) the Modified Minimum Non-locality Subtraction ($\overline{\text{MNS}}$) scheme, in analogy with the MS and $\overline{\text{MS}}$ schemes in DR.

\begin{figure}[!h] 
\centering
\includegraphics[width=0.3\textwidth]{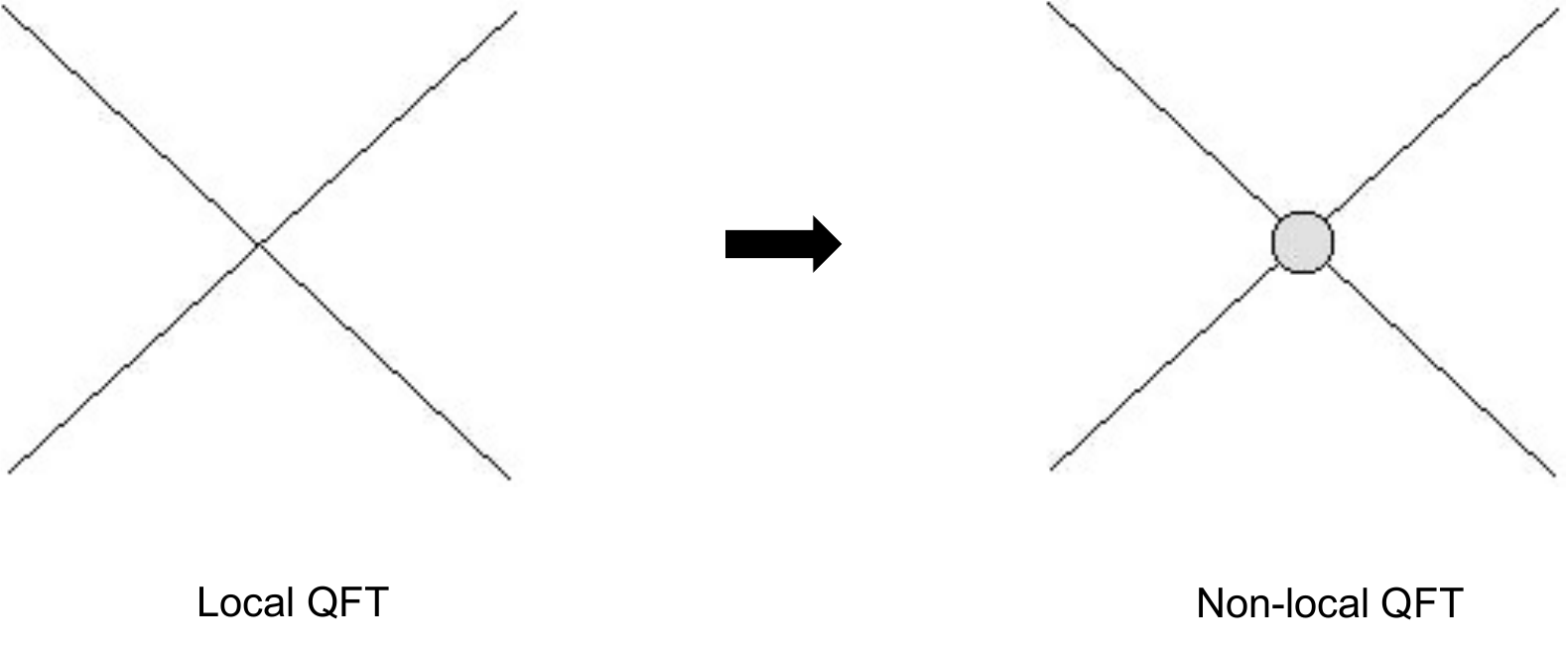}
\caption{Local (left) vs. non-local (right) interaction.}
\label{fig3}
\end{figure}

The physical interpretation of the NRS is as follows: non-locality can be viewed as the quantum fuzziness in the particle's wavefunction. It serves as to smear point-like interactions as in Fig. \ref{fig3}. This quantum fuzziness should be part of the definition of the renormalization of physical quantities and should not be physically observable, as it screens all scales smaller than it, in a manner similar to charge screening in local QED.

We illustrate the NRS with a few simply examples in the $\phi^{4}$ and $\phi^{3}$ theories.

\subsection{Renormalizing the $\phi^{4}$ Theory}\label{sec:phi4Ren}
First, let's consider the renormalization of the non-local $\phi^{4}$ theory given in eq. (\ref{eq:phi4Lag}). The quantities appearing in the Lagrangian are taken to be bare quantities. Thus, we first perform the following rescaling
\begin{equation}\label{eq:rescaling}
\phi_{0} = \sqrt{Z} \exp{\Big(\frac{m^{2}-m_{0}^{2}}{2\Lambda^{2}}\Big)}\phi,
\end{equation}
with the subscript indicating bare quantities. Next, we define the following conterterms
\begin{eqnarray}
 \delta_{Z} & = & Z - 1, \label{eq:counterterm1} \\ 
\delta_{m}& = &  Zm_{0}^{2} -  m^{2},  \label{eq:counterterm2}\\
\delta_{\lambda} & = & \lambda_{0} Z^{2} \exp{\Big(\frac{2 (m^{2} -m_{0}^{2})}{\Lambda^{2}}\Big)} - \lambda. \label{eq:counterterm3}
\end{eqnarray}

With these substitutions, eq. (\ref{eq:phi4Lag}) becomes
\begin{flalign}\label{eq:Renphi4Lag}
  \begin{aligned}
\mathscr{L}_{\phi} = & -\frac{1}{2}\phi e^{\frac{\Box +m^{2}}{\Lambda^{2}}}(\Box +m^{2})\phi - \frac{\lambda}{4!}\phi^{4} \\
& -\frac{1}{2}\phi e^{\frac{\Box +m^{2}}{\Lambda^{2}}}(\delta_{Z}\Box +\delta_{m})\phi - \frac{\delta_{\lambda}}{4!}\phi^{4},
\end{aligned}
\end{flalign}
with the $m$ and $\lambda$ now being the physical quantities. It is straightforward to obtain the Feynman rules of the non-local counterterms.
\begin{flalign}\label{eq:counterms}
\begin{aligned}    
 	\includegraphics[width=0.2\linewidth, valign=c,]{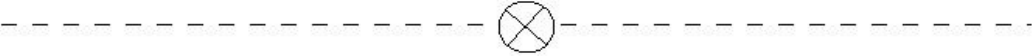}  & =   - i e^{\frac{-p^{2}+m^{2}}{\Lambda^{2}}}(-p^{2} \delta_{Z} + \delta_{m}), \\
 	\\
     \includegraphics[width=0.2\linewidth, valign=c,]{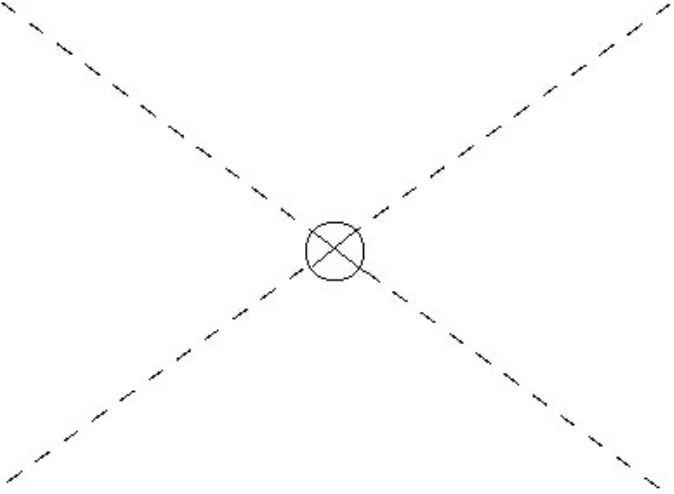}  & =  - i \delta_{\lambda}.
\end{aligned}
\end{flalign}  

With these counterterms, we can proceed with renormalizing "divergent" quantities. It is easy to show that in the $\phi^{4}$ theory, only the 2-point and 4-point functions need renormalization, as all other $n$-point functions are already finite in the limit $\Lambda \rightarrow \infty$.

\subsubsection{2-Point Function}
The non-renormalized non-local 2-point function is given in eq. (\ref{eq:2ptnon-local1}). In order to renormalize it, we add to it the first counterterm in eq. (\ref{eq:counterms}), and demand that sum be equal to the renormalized local result given in eq. (\ref{eq:2ptLocal2}) up to scheme-dependent constants. For example, in the MNS scheme, we only need to subtract $\Lambda^{2}$. Thus we impose 
\begin{equation}\label{eq:2ptMNS1}
-\frac{\lambda}{32\pi^{2}}e^{-\frac{m^{2}}{\Lambda^{2}}} \Lambda^{2} + e^{\frac{-p^{2}+m^{2}}{\Lambda^{2}}}\Big[ p^{2}\delta_{Z} -\delta_{m}\Big]_{p^{2}=m^{2}} = 0,
\end{equation}
where we have defined the renormalization on-shell for convenience. Eq. (\ref{eq:2ptMNS1}) does not specify the renormalization conditions uniquely, however, one can define the following suitable conditions
\begin{eqnarray}\label{eq:2pMNSconditions1}
\delta_{m} & = &  -\frac{\lambda}{32\pi^{2}}e^{-\frac{m^{2}}{\Lambda^{2}}} \Lambda^{2},\\
\delta_{Z} & = & 0,
\end{eqnarray}
which immediately define the renormalized 2-point function to be
\begin{equation}\label{eq:2ptMNS2}
\Pi_{2}^{\text{MNS}} = -\frac{\lambda m^{2}}{32\pi^{2}} \text{Ei} \Big(\frac{-m^{2}}{\Lambda^{2}} \Big).
\end{equation}

Notice that in the limit $\Lambda \rightarrow \infty$, we can expand the function $ \text{Ei} (-m^{2}/\Lambda^{2})$ as in eq. (\ref{eq:EiExpansion}). Keeping the leading term, we have
\begin{flalign}\label{eq:2ptMNS_matching}
\Pi_{2}^{\text{MNS}}  & \simeq -\frac{\lambda m^{2}}{32\pi^{2}}  \Big[ \gamma_{E} + \log{\Big( \frac{m^{2}}{\Lambda^{2}}\Big)} \Big]\nonumber\\
& = \Pi_{2}^{\text{Local}}(\mu^{2}  = \Lambda^{2}) + \text{constant terms},
\end{flalign}
as it should. On the other hand, in $\overline{\text{MNS}}$, we need to subtract the divergence in $\text{Ei}(-m^{2}/\Lambda^{2})$ in addition to subtracting $\Lambda^{2}$. Therefore, instead of setting the L.H.S of eq. (\ref{eq:2ptMNS1}) to vanish, we set it equal to
\begin{equation}\label{eq:2ptMNSbar1}
\frac{\lambda m^{2}}{32\pi^{2}} \Big[1 - \text{Ei}\Big(\frac{-m^{2}}{\Lambda^{2}} \Big) + \text{Ei}\Big(\frac{-\mu^{2}}{\Lambda^{2}} \Big) \Big].
\end{equation}

Imposing this renormalization condition with the on-shell condition assumed, it is fairly simple to see that $\delta_{m}$ remains unchanged, whereas $\delta_{Z}$ becomes
\begin{equation}\label{eq:2pMNSbarconditions1}
\delta_{Z}= \frac{\lambda}{32\pi^{2}} \Big[1 + \text{Ei}\Big(\frac{-\mu^{2}}{\Lambda^{2}} \Big)\Big],
\end{equation}
and the renormalized 2-point function in the $\overline{\text{MNS}}$ reads
\begin{equation}\label{eq:2ptMNSbar2}
\Pi_{2}^{\overline{\text{MNS}}} = \frac{\lambda m^{2}}{32\pi^{2}} \Big[1+  \text{Ei} \Big(\frac{-\mu^{2}}{\Lambda^{2}} \Big) - \text{Ei} \Big(\frac{-m^{2}}{\Lambda^{2}} \Big) \Big],
\end{equation}
and it is a simple exercise to verify that it reduces to the renormalized local case given in eq. (\ref{eq:2ptLocal2}) in the limit $\Lambda \rightarrow \infty$.

\subsubsection{4-Point Function}
Now we turn our attention to renormalizing the 4-point function. Notice that the interaction and its counterterm (second line in eq. (\ref{eq:counterms})) are identical to the local case, however, the propagators are modified by the non-locality form factor, which means that the effect of non-locality only arises from loops. The contributions to the 4-point functions are shown in Fig. \ref{fig4} below.
\begin{figure}[!h] 
\centering
\includegraphics[width=0.45\textwidth]{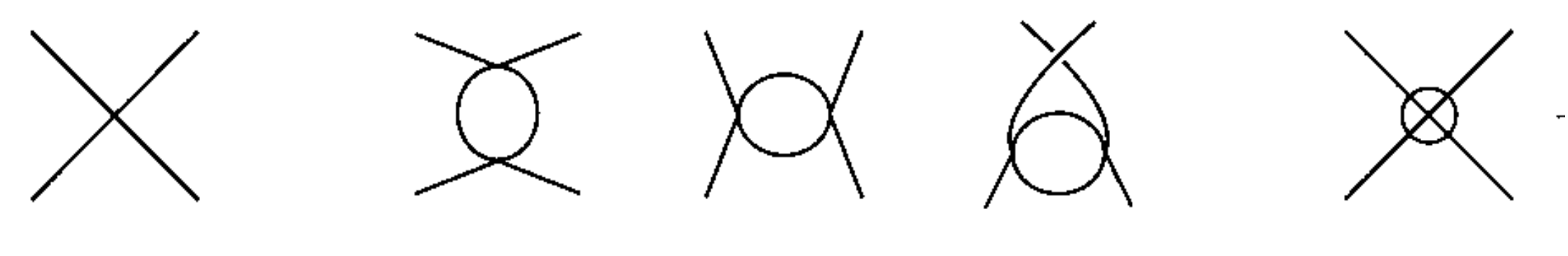}
\caption{Contributions to the $\phi^{4}$ 4-point function.}
\label{fig4}
\end{figure}

Following \cite{Peskin:1995ev} for the local case, the renormalization condition is set to be $i\Gamma_{4} = -i\lambda$ at $s = 4m^{2}$ and $t = u =0$, with $s, t$ and $u$ being the Mandelstam variables. The renormalized local 4-point function thus reads
\begin{flalign}\label{eq:4ptLocalCondition}
  \begin{aligned}
\Gamma_{4}^{\text{Local}} & =  -\lambda - \frac{\lambda^{2}}{32\pi^{2}}\int_{0}^{1}dx \Big[ \log{\Big( \frac{f(s)}{f(4m^{2})}\Big)} \\
 & + \log{\Big( \frac{f(t)}{f(0)}\Big)} + \log{\Big( \frac{f(u)}{f(0)}\Big)} \Big].],
	\end{aligned}
\end{flalign}
where $f(\tau) = (m^{2}-x(1-x)\tau)$. On the other hand, the non-local matrix element reads
\begin{equation}\label{eq:4ptNL}
\Gamma_{4}^{\text{NL}} = -\lambda+(-i\lambda)^{2}\Big[V(s)+V(t) + V(u) \Big] - \delta_{\lambda},
\end{equation}
where
\begin{equation}\label{eq:nlV}iV(p^{2}) = \frac{1}{S}\int \frac{d^{4}k}{(2\pi)^{4}} \frac{ie^{\frac{k^{2}-m^{2}}{\Lambda^{2}}}}{k^{2}-m^{2}} \frac{ie^{\frac{(p+k)^{2}-m^{2}}{\Lambda^{2}}}}{(p+k)^{2}-m^{2}},
\end{equation}
where $S=2$ is a symmetry factor. In general, the momentum integral is not doable exactly, however, in the limit $\Lambda^{2} \gg m^{2}, p^{2}$, it can be done to a good level of approximation. We show how to evaluate this integral approximately in the Appendix. Given the approximate result in eq. (\ref{eq:NL4pt4}), we see that the non-local 4-point function lacks a $\Lambda^{2}$ term, and thus it is already renormalized in the MNS scheme, however, it does contain the term $\text{Ei}(-2m^{2}/\Lambda^{2})$, which still needs to be subtracted in the $\overline{\text{MNS}}$ scheme. So, plugging eq. (\ref{eq:NL4pt4}) in eq. (\ref{eq:4ptNL}) and imposing the same renormalization condition as in the local case,	 we arrive at the following renormalization condition
\begin{equation}\label{eq:4ptNLcondition}
\delta_{\lambda} \simeq -\frac{\lambda^{2}}{32\pi^{2}}e^{-\frac{2m^{2}}{\Lambda^{2}}} \Big[1 + 3\text{Ei} \Big(\frac{-2m^{2}}{\Lambda^{2}}\Big)\Big],
\end{equation}
and the renormalized non-local 4-point function immediately follows
\begin{flalign}\label{eq:renormalizedNL4pt}
\begin{aligned}
\Gamma_{4}^{\overline{\text{MNS}}} &  \simeq -\lambda -\frac{\lambda^{2}}{32\pi^{2}}e^{-\frac{2m^{2}}{\Lambda^{2}}} \int_{0}^{1} dx \Bigg[2 - 3\text{Ei}\Big(\frac{-2m^{2}}{\Lambda^{2}} \Big) \\
& + \text{Ei}\Big(\frac{-2f(s)}{\Lambda^{2}} \Big)  + \text{Ei}\Big(\frac{-2f(t)}{\Lambda^{2}} \Big)  + \text{Ei}\Big(\frac{-2f(u)}{\Lambda^{2}} \Big)   \Bigg].
\end{aligned}
\end{flalign}

Expanding the exponential integral functions and keeping in mind that
\begin{equation}\label{eq:trick}
-\int_{0}^{1}\log{(1-2x)^{2}} = 2,
\end{equation}
it is easy to show that $\Gamma_{4}^{\overline{\text{MNS}}}$ reduces to the local case in eq. (\ref{eq:4ptLocalCondition}) in the limit $\Lambda \rightarrow \infty$.

\subsection{Renormalizing the $\phi^{3}$ Theory}\label{sec:phi3Ren}
Another example that will be useful for our purposes is the non-local $\phi^{3}$ theory, which is essentially given by eq. (\ref{eq:phi4Lag}) with the interaction term being replaced with $\frac{1}{3!}\kappa_{0}\phi^{3}$. Performing the rescaling in eq. (\ref{eq:rescaling}), and defining the counterterms in eqs. (\ref{eq:counterterm1}) and  (\ref{eq:counterterm2}), in addition to the following conterterm for $\kappa_{0}$ 
\begin{equation}\label{eq:counterterm4}
	\delta_{\kappa} = \kappa_{0} Z\sqrt{Z} \exp{\Big(\frac{3 (m^{2} -m_{0}^{2})}{2\Lambda^{2}}\Big)} - \kappa,
\end{equation}
it is easy to see that the propagator counterterm remains unchanged, whereas the interaction counterterm is simply given by $-i\delta_{\kappa}$.

It is fairly simple to show that all $n$-point functions for $n > 2$ in the non-local theory are finite in the limit $\Lambda \rightarrow \infty$, and therefore they do not need to be renormalized. This leaves us with only the 2-point function, which naively has a logarithmic divergence (in the limit $\Lambda \rightarrow \infty$). 

Starting with the local theory, it is fairly simple to show that in the $\overline{\text{MS}}$ scheme, the renormalized 2-point function reads
\begin{equation}\label{eq:2ptPhi3Local}
	\Pi_{2}^{\text{Local}} = -\frac{\kappa^{2}}{32\pi^{2}} \Big[\log{\Big(\frac{m^{2}}{\mu^{2}}\Big)} + C\Big],
\end{equation}
where we have collected all constant terms in $C$. Following the $\overline{\text{MNS}}$ prescription, we can define the renormalization conditions as 
\begin{eqnarray}\label{eq:2pMNSbarconditions2}
\delta_{m} & =&    \frac{-\kappa^{2}e^{\frac{-2m^{2}}{\Lambda^{2}}}}{32\pi^{2}} \int_{0}^{1} dx \Big\{ 1 - C + \text{Ei}\Big[\frac{g(x)\mu^{2}}{\Lambda^{2}} \Big] \Big\}\\
\delta_{Z} & =&   0,
\end{eqnarray}
where $g(x) = -2(1-x+x^{2})$ and thus the renormalized 2-point function reads
\begin{flalign}\label{eq:Ren2ptphi3}
\Pi_{2}^{\overline{\text{MNS}}} & = \frac{\kappa^{2}}{32\pi^{2}}e^{\frac{-2m^{2}}{\Lambda^{2}}} \int_{0}^{1} dx \Big\{\text{Ei}\Big[\frac{g(x)\mu^{2}}{\Lambda^{2}}\Big] \nonumber \\
& - \text{Ei}\Big[\frac{g(x) m^{2}}{\Lambda^{2}}\Big] -C \Big\},
\end{flalign}
which can be shown to reduce to eq. (\ref{eq:2ptPhi3Local}) in the limit $\Lambda \rightarrow \infty$ by simply expanding the exponential integral function.

\subsection{The $\phi^{4}$ $\beta$-Function}
Since the $\beta$-function should be independent of the renormalization scheme, it is worthwhile to compare the $\beta$-function in both the renormalized and unrenormalized cases and see if they agree. This would provide a good sanity check for the NRS. Here we do it for the $\phi^{4}$ theory, and to keep things simple, we consider the massless case. The 1-loop $\beta$-function of the unrenormalized case is given by \cite{Ghoshal:2017egr}
\begin{equation}\label{eq:betaFunc1}
\beta_{\lambda}^{(1)} = \frac{3\lambda^{3}}{16\pi^{2}}e^{-\frac{2\mu^{2}}{\Lambda^{2}}}.
\end{equation}

The $\beta$-function can be found from the 4-point function calculated with a UV-cutoff $M$ as follows
\begin{equation}\label{eq:betaFunc2}
\beta_{\lambda}^{(1)}  = - M \frac{\partial}{\partial M}\Gamma_{4}\Bigg|_{M \rightarrow \mu},
\end{equation}
where $\mu$ is the renormalization scale. In the massless limit, the 4-point function reads (see eq. (\ref{eq:4ptNL}))
\begin{equation}\label{eq:4-pt_massless}
\Gamma_{4} = -\lambda - 3\lambda^{2}V(p^{2}=0) -\delta_{\lambda},
\end{equation}
where $V(p^{2})$ is given by eq. (\ref{eq:nlV}) with $m^{2} \rightarrow 0$. Requiring that the 4-point function be equal to $-i\lambda$ when $s=t=u=0$ fixes the counterterm to be
\begin{equation}\label{eq:masslessCT}
\delta_{\lambda} = -3 \lambda^{2} V(p^{2}=0) = \frac{3\lambda^{2}}{2} \int^{M}\frac{d^{4}k}{(2\pi)^{4}}\frac{e^{2k^{2}/\Lambda^{2}}}{k^{4}}.
\end{equation}

It is easy to see that the only dependence on the UV cutoff in the 4-point function arises from the counterterm. Thus, the $\beta$ function becomes
\begin{equation}\label{eq:betaFunc1}
\beta_{\lambda}^{\text{NSR}(1)} =  M \frac{\partial \delta_{\lambda}}{\partial M} \Bigg|_{M \rightarrow \mu} = \frac{3\lambda^{2}}{16\pi^{2}}e^{-\frac{2\mu^{2}}{\Lambda}}, 
\end{equation}
which agrees with eq. (\ref{eq:betaFunc1}), thereby exonerating the NRS.

\section{Application to the Hierarchy Problem}\label{sec:HierarchyProblem}
\subsection{The Old View and the Modern View}\label{sec:modern}
The old view of the hierarchy problem is related to the issue of the quadratic (and to a lesser extent logarithmic) divergences in the corrections to the Higgs mass. More specifically, when calculating the 1-loop corrections to the Higgs mass using a UV cut-off $\Lambda_{UV}$, these corrections turn out to be proportional to $\Lambda_{UV}^{2}$ and/or $\log{(\Lambda_{UV}^{2}/m_{}^{2})}$, thus rendering the Higgs mass unnatural as the UV cutoff is taken to be large, and avoiding this would require unnatural fine-tuning \cite{Susskind:1978ms, Susskind:1982mw}.

With the old view of the hierarchy problem, the potential of non-local QFTs to solve these divergences was recognized early on, because non-local momentum integrals are finite in the limit $k \rightarrow \infty$, since they have the scale of non-locality as a natural cutoff that regularizes them (see for instance \cite{Krasnikov:1987yj, Moffat:1988zt, Moffat:1990jj}). However, the price that one pays is that instead of the quadratic divergence, the mass corrections become quadratically dependent on the scale of non-locality as we have shown earlier. This meant that in order to have a natural solution that is not fine-tuned, the scale of non-locality had to be low, potentially $\Lambda \sim 1-10$ TeV.

The null results in the LHC (albeit still leaving some room for new physics below $\sim 10$ TeV), have induced a different view of the hierarchy problem. One no longer refers to any quadratic divergences, as DR is used to regularize divergent integrals of the Higgs mass corrections arising from the SM degrees of freedom. However, the Higgs sector would still suffer from a naturalness problem if it coupled to some heavy sector, as this sector would yield large threshold corrections even after subtracting the divergences with DR. To better illustrate the modern viewpoint of the hierarchy problem, let's assume that the SM Higgs couples to some heavy particle of mass $M \gg m_{h}$. Viewing the SM as an EFT of the full UV sector, one can show that once this UV sector has been integrated out, the physical Higgs mass would schematically be given by
\begin{equation}\label{eq:HiggsEFT}
m_{h}^{2} = m_{H}^{2} - \frac{1}{16\pi^{2}}\Big[C_{0}M^{2} + C_{1} m_{H}^{2} + C_{2} \frac{m_{H}^{4}}{M^{2}} + \cdots \Big],
\end{equation}
where $m_{H}$ is the bare mass, and $C_{i}$ are Wilson coefficients arising from integrating out higher-order operators that contribute to the Higgs mass from the heavy sector. One can see that we need to tune $\frac{M}{4\pi} \gg m_{h}$ against $m_{H}$ in order to have $m_{h} \sim O(100)$ GeV, which is obviously unnatural. This is precisely the origin of the fine-tuning problem in the Higgs sector, which is most transparent in this Wilsonian approach.\footnote{We refer the interested reader to \cite{Branchina:2022jqc} for an in-depth discussion of the hierarchy problem both from the Wilsonian view and the DR view, where it is shown that these two approaches agree.} On the other hand, if the Higgs boson does not couple to any UV sector, then its mass will not receive any large threshold corrections and the fine-tuning problem is averted. Nonetheless, there are compelling reasons to believe that the Higgs does indeed couple to a UV sector (or sectors), most notable of which are to provide a natural explanation of neutrino masses via the see-saw mechanism, the unification of gauge couplings at the GUT scale, and the emergence of quantum gravity at the Planck scale. Thus, the hierarchy problem is indeed worth careful investigation.

\subsection{Non-locality as a Solution in the Modern View}\label{sec:NLsol}
Armed with the NRS, we would like to evaluate the viability of (renormalized) non-local QFTs for solving the hierarchy problem in its modern form. To this avail, we shall analyze a simplified scalar toy model that consists of a light Higgs-like scalar singlet $\phi$ of mass $ m \sim O(\text{EW})$ that is coupled to another heavy scalar singlet $H$ with mass $M \gg m$. To keep our analysis simple, we shall impose $Z_{2}$ symmetry on $\phi$ and assume a universal non-locality scale. The most general Lagrangian that one can write is
\begin{flalign}\label{eq:UVLag}
\mathscr{L}_{\phi H} & = -\frac{1}{2}\phi e^{\frac{\Box +m^{2}}{\Lambda^{2}}}(\Box +m^{2})\phi  -\frac{1}{2}H e^{\frac{\Box +M^{2}}{\Lambda^{2}}}(\Box +M^{2})H \nonumber\\ & - \frac{\lambda_{1}}{4!}\phi^{4} -\frac{\lambda_{2}}{3!}M H^{3}-\frac{\lambda_{3}}{2}M \phi^{2}H-\frac{\lambda_{4}}{4}\phi^{2}H^{2},
\end{flalign}
where we have factored out the mass of $H$ in the trilinear couplings. In general, we could also have a $H^{4}$ term, however, it is irrelevant for our analysis.
 
\begin{figure}[!t] 
\centering
\includegraphics[width=0.25\textwidth]{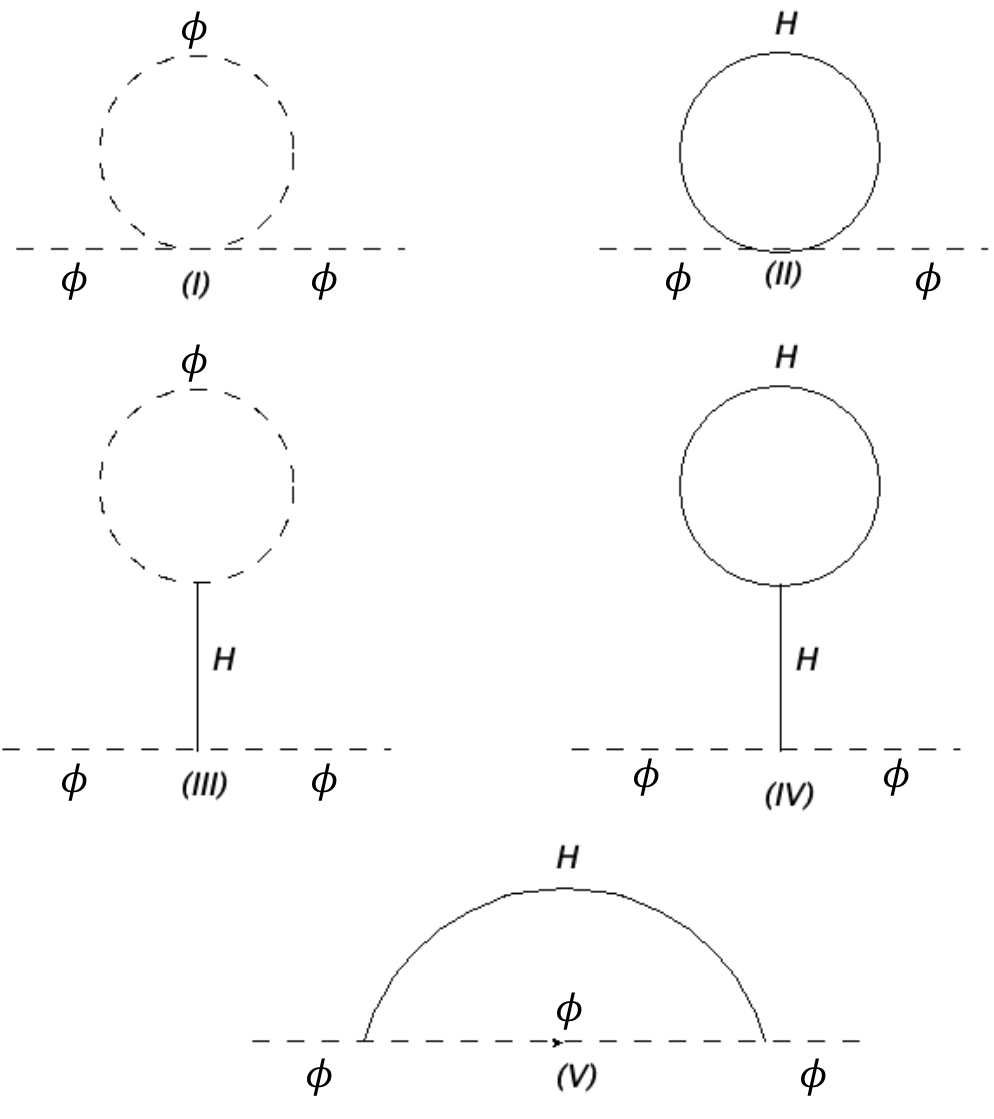}
\caption{$1$-loop corrections to the mass of $\phi$.}
\label{fig5}
\end{figure}

The 1-loop corrections of the mass of $\phi$ are shown in Fig. \ref{fig5}. In calculating the mass corrections, it is convenient to employ the MNS scheme. But before doing so, let's discuss the unrenormalized case. The unsubtracted corrections are given by
\begin{eqnarray}
\delta m_{\text{1}}^{2} & = &  \frac{\lambda_{1}}{32\pi^{2}} e^{-\frac{m^{2}}{\Lambda^{2}}} \Big[ \Lambda^{2} + m^{2} e^{\frac{m^{2}}{\Lambda^{2}}}\text{Ei}\Big(\frac{-m^{2}}{\Lambda^{2}}\Big)\Big], \label{eq:mCorr1}\\
\delta m_{\text{2}}^{2}  & = &  \frac{\lambda_{4}}{32\pi^{2}}e^{-\frac{M^{2}}{\Lambda^{2}}}\Big[\Lambda^{2} + M^{2} e^{\frac{M^{2}}{\Lambda^{2}}}\text{Ei}\Big(\frac{-M^{2}}{\Lambda^{2}}\Big)\Big], \label{eq:mCorr2}\\
\delta m_{\text{3}}^{2}  & = &\frac{-\lambda_{3}^{2}}{32\pi^{2}}e^{\frac{-m^{2}-M^{2}}{\Lambda^{2}}} \Big[ \Lambda^{2} + m^{2} e^{\frac{m^{2}}{\Lambda^{2}}} \text{Ei}\Big(\frac{-m^{2}}{\Lambda^{2}}\Big) \Big], \label{eq:mCorr3}\\
\delta m_{\text{4}}^{2}  & = & - \frac{\lambda_{2}\lambda_{3}}{32\pi^{2}} e^{-\frac{2M^{2}}{\Lambda^{2}}} \Big[ \Lambda^{2} + M^{2}e^{\frac{M^{2}}{\Lambda^{2}}} \text{Ei}\Big( \frac{-M^{2}}{\Lambda^{2}}\Big) \Big], \label{eq:mCorr4}\\
\delta m_{\text{5}}^{2}  & = & \frac{\lambda_{3}^{2}}{32\pi^{2}}M^{2}e^{\frac{M^{2}}{\Lambda^{2}}}  \text{Ei}\Big(\frac{-2M^{2}}{\Lambda^{2}}\Big) \label{eq:mCorr5},
\end{eqnarray}
where the MNS renormalized corrections are obtained simply by dropping the $\Lambda^{2}$ term from  eqs. (\ref{eq:mCorr1})-(\ref{eq:mCorr4}) with eq. (\ref{eq:mCorr5}) remaining unchanged. Notice that $\delta m_{2-5}^{2}$ arise from the UV sector, whereas $\delta m_{1}^{2}$ is an IR threshold correction due to the light scalar's self-energy.

In the unrenormalized case, and for $\Lambda \gtrsim M$, all corrections in eqs. (\ref{eq:mCorr1})-(\ref{eq:mCorr5}) are too large, i.e. non-locality does not protect the mass $m$ from any large corrections from the heavy sector. This is easy to understand by inspecting the exponential form factor used to express non-locality. On the other hand, when $\Lambda$ is \textit{sufficiently smaller} than $M$, then it is easy to show that the  corrections (\ref{eq:mCorr2})-(\ref{eq:mCorr5}) are always exponentially suppressed, and thus non-locality does indeed protect the mass from receiving large corrections from the UV sector, however, the IR contribution (\ref{eq:mCorr1}) will now acquire a threshold correction $\sim \Lambda^{2}$, which will be large unless $\Lambda$ is not much larger than $m$. 

\begin{figure}[!h] 
\centering
\includegraphics[width=0.35\textwidth]{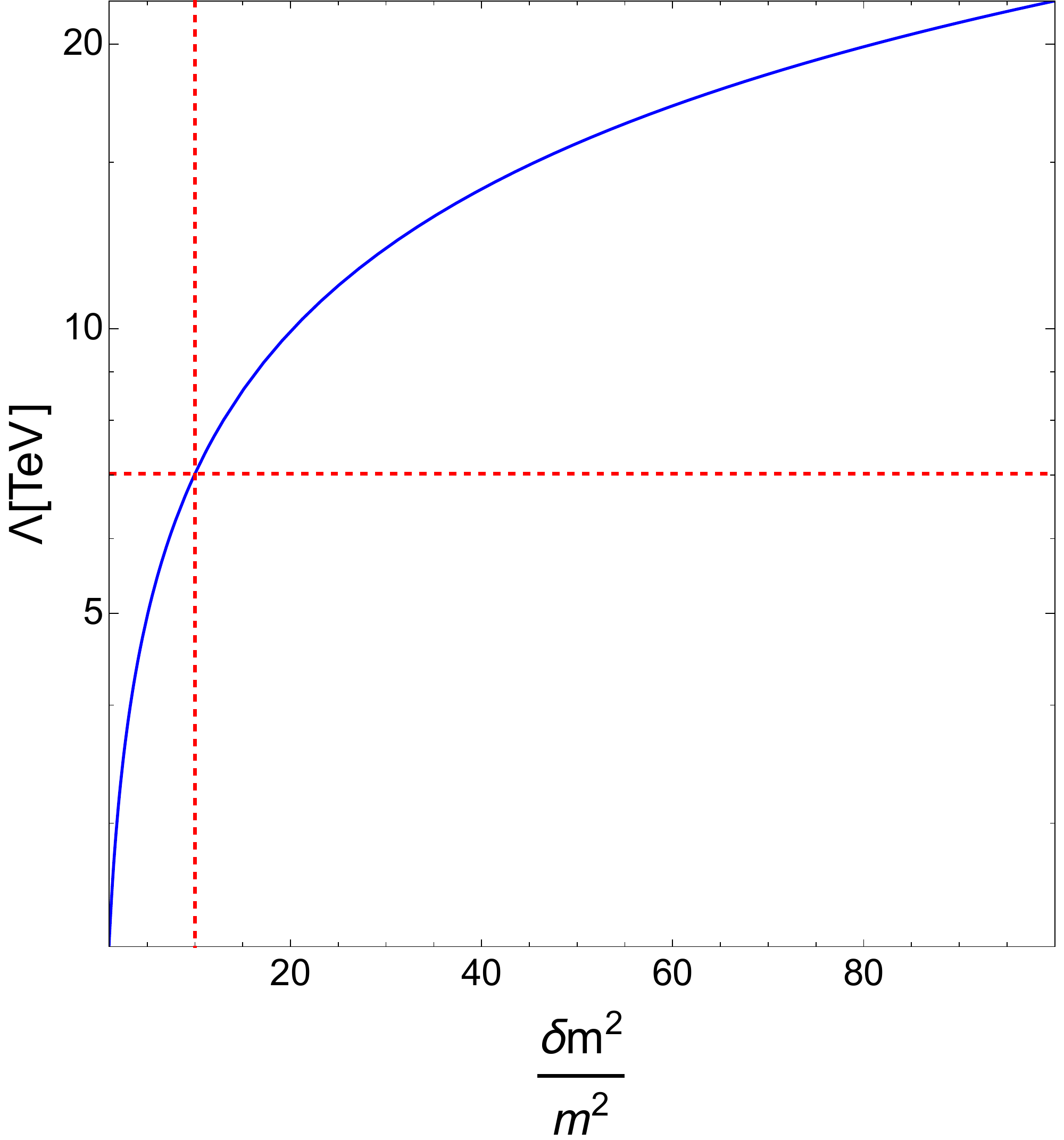}
\caption{The scale non-locality $\Lambda$ plotted against the level of fine-tuning in the mass of the light scalar $\phi$. Here we set $\lambda_{1} = 1$ and $m = 125$ GeV. The red dashed lines indicate a fine-tuning level of $\delta m^{2}/m^{2} = 10\%$ and the corresponding scale of $7$ TeV. Notice that this is the IR correction only as all corrections from the UV sector are exponentially suppressed for $\Lambda$ sufficiently smaller than $M$.}
\label{fig6}
\end{figure}

Despite the threshold correction in the self-energy of $\phi$, it is possible to keep the mass of $\phi$ naturally light while at the same time evade the constraints from the LHC. If we estimate the level of fine-tuning in the mass of $\phi$ as $\delta m^{2}/m^{2}$, then evading the LHC limit of $\Lambda \gtrsim 3$ TeV \cite{Biswas:2014yia}, would require fine-tuning of only $\sim O(50\%)$. On the other hand, if we demand that the level of fine-tuning not exceed $ O(10\%)$, then we must have $\Lambda \lesssim 7$ TeV. We show the scale of non-locality as a function of the fine-tuning in Fig. \ref{fig6}.

Notice that the (renormalized) local case in eq. (\ref{eq:2ptLocal2}) yields smaller corrections $\sim \frac{m^{2}}{32\pi^{2}}$ as opposed to  $\sim \frac{\Lambda^{2}}{32\pi^{2}}$. So, when introducing non-locality without renormalization, one loses in the IR sector what one gains in the UV sector.

\begin{figure}[!t] 
\centering
\includegraphics[width=0.35\textwidth]{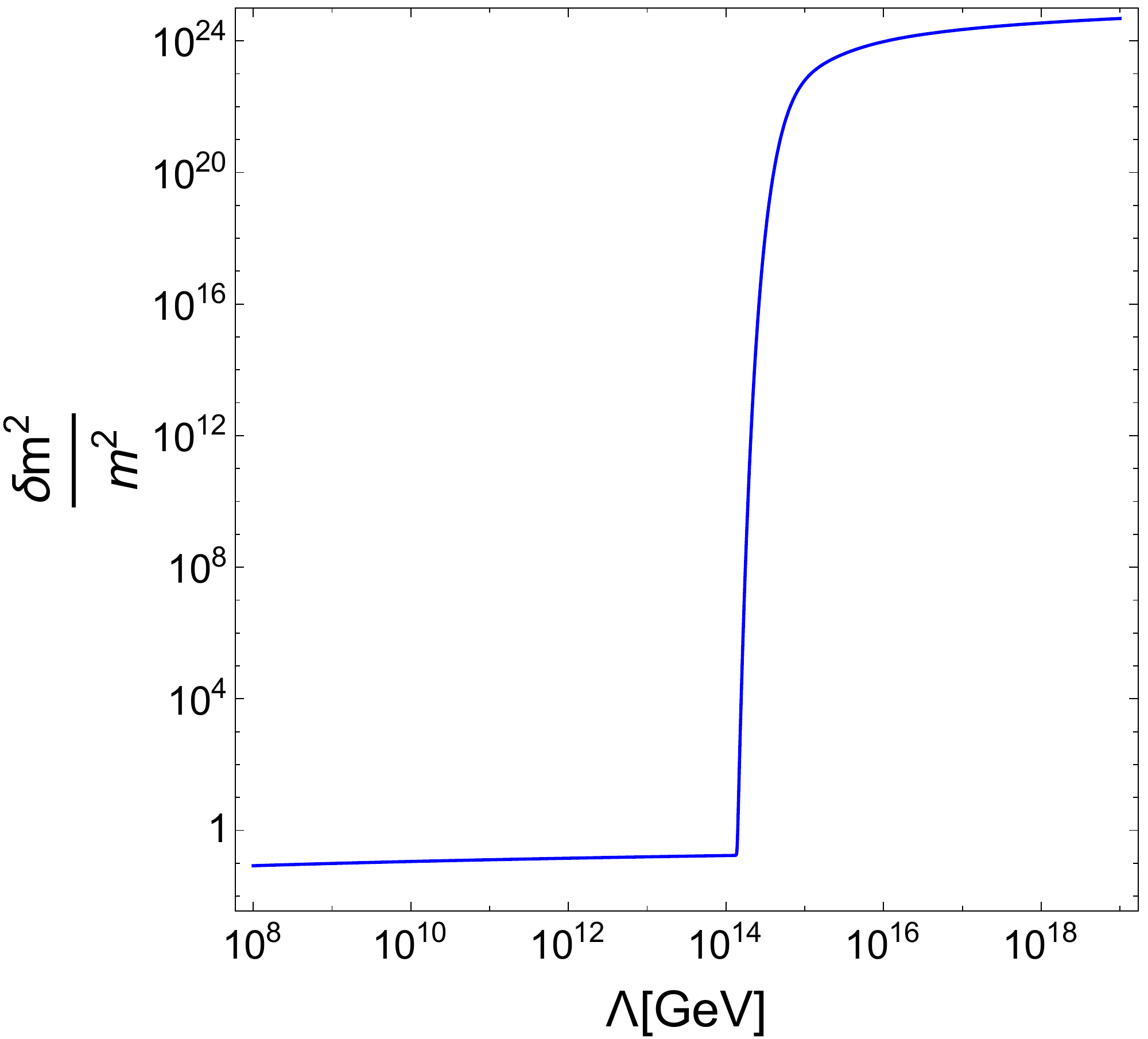}
\caption{The MNS-subtracted mass corrections as a function of the non-locality scale $\Lambda$. Here, we fix $\lambda_{1} = 1$, $\lambda_{2}=0.8$, $\lambda_{3}=1.2$, $\lambda_{4}=0.9$, $m=125$ GeV and $M = 10^{15}$ GeV.}
\label{fig7}
\end{figure}

As we argued earlier, this is merely an artifact of not renormalizing the non-local theory, and things improve drastically when we apply the NRS, as large IR corrections are completely absent. To be more specific, in the MNS scheme, all corrections in eqs. (\ref{eq:mCorr1})-(\ref{eq:mCorr5}) will be suppressed as long as $\Lambda$ is sufficiently smaller than $M$, i.e. not only is the mass of $\phi$ protected from large corrections from the UV sector, but also from the threshold corrections of the IR sector, consistent with the dimensionally-regularized local case. We show this suppression in Fig. \ref{fig7}. As the figure shows, all mass corrections remain negligible as long as $\Lambda$ is sufficiently smaller than $M$. However, once $\Lambda$ becomes sufficiently close to $M$, the corrections from the UV sector starts to increase sharply and eventually saturate to a very large value $\sim M$, similar to the unsubtracted case. This behavior can easily be understood by inspecting eqs. (\ref{eq:mCorr2}), (\ref{eq:mCorr4}) and (\ref{eq:mCorr5}). After subtracting $\Lambda^{2}$ in the MNS scheme, we can see that all of these corrections are proportional to $M^{2}\text{Ei}(-M^{2}/\Lambda^{2})$, which becomes vanishingly small in the limit $M/\Lambda \gg 1$, because $\lim_{x \rightarrow \infty} \text{Ei}(-x) =0$. On the other hand, when $M/\Lambda \ll 1$, $M^{2}\text{Ei}(-M^{2}/\Lambda^{2}) \simeq M^{2}\log{(-M^{2}/\Lambda^{2})} + O(M^{4}/\Lambda^{2})$ yielding large corrections. The situation isn't improved by employing the $\overline{\text{MNS}}$ scheme, as the corrections become $\sim M^{2}[\text{Ei}(-M^{2}/\Lambda^{2}) - \text{Ei}(-\mu^{2}/\Lambda^{2})]$, which would yield large corrections unless we set $\mu^{2} \sim M^{2}$, and the result isn't much different from the MNS scheme.

We can therefore summarize our main conclusion as follows: Non-locality eliminates all mass corrections from all UV sectors at mass scales that are sufficiently larger than the scale of non-locality, and applying NRS obviates any large threshold corrections from the IR sector that emanate from introducing the said non-locality, in addition to improving the UV behavior of the non-local theory, however, non-locality does not protect the mass of the any scalar in the IR sector from receiving large corrections from UV sectors at mass scales comparable to or smaller than $\Lambda$. Therefore, in order to furnish an acceptable solution to the hierarchy problem that avoids fine-tuning, we need to assume that the SM Higss does not couple to any sector in the region $\sim O(\text{EW}) \ll E \sim \Lambda$, whereas it can couple to any other UV sectors at scales sufficiently larger than $\Lambda$. 

This assumption is not as constraining as it seems, since the scales of NP to which the Higgs is expected to couple are the GUT scale and and Plank scale. Therefore, if we assume $\Lambda$ to be sufficiently smaller than the GUT scale, all the problematic corrections to its mass are filtered out. This also allows us to push the scale of non-locality to very large scales,\footnote{In \cite{Abu-Ajamieh:2021vnh}, unitarity was used to derive a model-independent Veltman condition, which yielded an upper bound of $\sim 19$ TeV on the scale of any NP that can solve the hierarchy problem. However, that analysis was conducted in the local case, and calculating the loop corrections in the renormalized non-local theory would avoid this upper limit} which would both help evade all experimental constraints, and confine any acausality to very short distances, rendering its effects highly suppressed at lower energies. Furthermore, non-locality can allow supersymmetry (SUSY) to be realized at much higher energies while still being natural, as long as the SUSY-breaking scale is sufficiently larger than the scale of nonlocality to eliminate threshold corrections from superpartners.
 
\begin{figure}[!t] 
\centering
\includegraphics[width=0.37\textwidth]{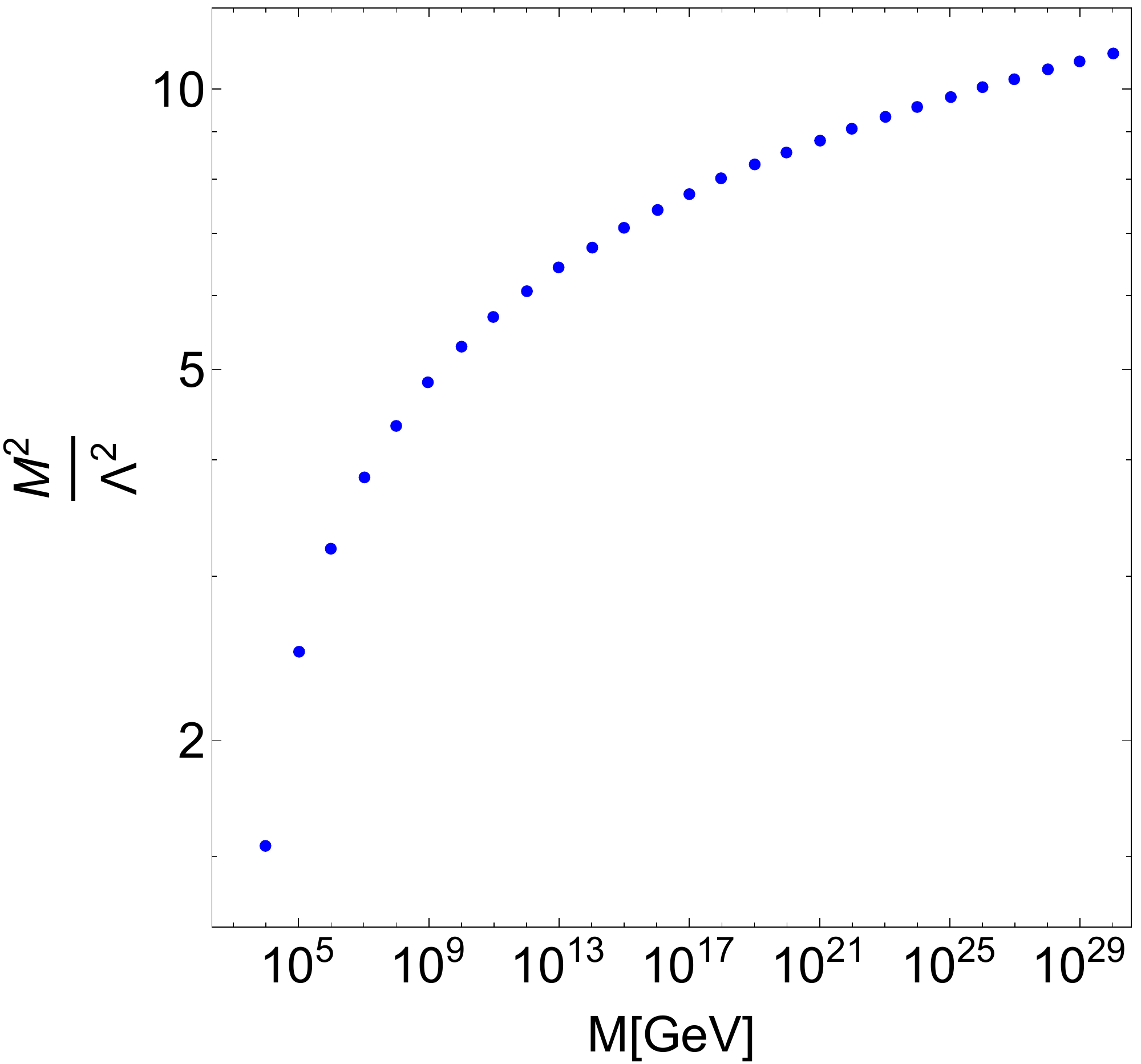}
\caption{The ratio $\frac{\Lambda^{2}}{M^{2}}$ needed to keep $\frac{\delta m^{2}}{m^{2}}=1$ as a function of the mass of the heavy sector $M$. The benchmark values are the same as in Fig. \ref{fig6}. The plot shows that as we increase the scale of the UV sector, the scale of non-locality needs to kick in earlier. Nonetheless, the dependence is logarithmic, and $\Lambda$ that is roughly an order of magnitude smaller than $M$ is sufficient to suppress all contributions from the UV sector.}
\label{fig8}
\end{figure}

Before we conclude this section, we need to investigate how "sufficiently smaller" $ \Lambda$ needs to be compared to $M$. As Fig. \ref{fig7} shows, the mass corrections start becoming large at a scale slightly smaller than $M$. To quantify this, we plot in Fig. \ref{fig8} the ratio $\frac{\Lambda^{2}}{M^{2}}$ at which the tuning $\frac{\delta m^{2}}{m^{2}}=1$, against $M$. As the plot shows, one must increase the separation between the non-locality scale and the scale of the UV sector as the latter increases. This dependence is logarithmic. We can see from the plot that even for very large scales of the UV sector $M$, the scale of non-locality should be smaller by roughly an order of magnitude at most in order to provide the required suppression.

\section{Conclusions}\label{Sec:conclusions}
In this paper, we formulated a renormalization scheme inspired by DR for non-local QFTs with infinite derivatives. In this scheme which we called NRS, we defined two subtraction schemes to eliminate the quadratic dependence (MNS) and the logarithmic-like dependence ($\overline{\text{MNS}}$) on the non-locality scale, which are divergent in the limit $\Lambda \rightarrow \infty$. We showed how renormalizing non-local QFTs links with the dimensionally renormalized local case and we illustrated this scheme using a few simple examples in the $\phi^{3}$ and $\phi^{4}$ theories.

We applied our results to the hierarchy problem in its modern view using a simple toy model, and we showed that when we enhance non-local QFTs with NRS, then not only does non-locality protect the mass of light scalars from large corrections arising from any UV sector that is sufficiently larger than the scale of non-locality, but also it can eliminate all threshold corrections from the IR sector. However, we found that non-locality does not eliminate large corrections to the mass of the scalar from heavy sectors that reside at scales comparable to or smaller than $\Lambda$. This implies that for the non-locality of be a viable solution to the SM hierarchy problem, then the SM Higgs can only couple to UV sectors sufficiently larger than the scale of non-locality.

Although we have formulated our scheme using a specific non-local form factor that is the exponential of an entire function, generalizing the NRS to any non-local form factor that is an entire function should be straightforward. It would also be interesting to apply the NRS to other non-local sectors, such as non-local QED and the non-local EW sector and investigate their applications to Electroweak Precision Observables (EWPO), especially the STU parameters, in order to set limits on the scale of non-locality. We postpone this to future work.

\section*{Acknowledgments}
FA thanks Sachin Vaidya for the valuable discussion, and Justin David and Nobuchika Okada for reading and commenting on the manuscript. This work was supported by the C.V. Raman fellowship from CHEP at IISc. 

\appendix

\section{4-Point Scalar Integral}
Here we show how to approximate the 4-point scalar integral given in eq. (\ref{eq:nlV}) in the limit $m^{2}, p^{2} \ll \Lambda^{2}$. In general, this limit is quite sufficient for $m^{2}, p^{2} \sim O(\text{EW})$ in light of the LHC constraints on the scale of non-locality.

Introducing the Feynman parameter and shifting $k$, it is easy to show that eq. (\ref{eq:nlV}) can be written as
\begin{equation}\label{eq:NL4pt1}
iV(p^{2})  = -\frac{1}{2} e^{-\frac{2m^{2}}{\Lambda^{2}}} \int_{0}^{1} dx \int \frac{d^{4}k}{(2\pi)^{4}} \frac{\exp{\Big[ \frac{2k^{2}-2 a p.k + b p^{2}}{\Lambda^{2}}  \Big]}}{(k^{2}-\Delta)^{2}},
\end{equation}
where $a=(1-2x)$, $b=(1-2x+2x^{2})$ and $\Delta = m^{2}+ x(x-1)p^{2}$. Now, expanding the linear term in $k$ in the exponent
\begin{equation}\label{eq:expansion1}
\exp{\Big(\frac{-2a p.k}{\Lambda^{2}}\Big)} = 1- \Big( \frac{2a p.k}{\Lambda^{2}}\Big) + \frac{1}{2!} \Big(\frac{2ap.k}{\Lambda^{2}}\Big)^{2} + \cdots,
\end{equation}
and noting that the function $e^{2k^{2}}/(k^{2}-\Delta)^{2}$ is even in $k$, it follows that all odd terms in eq. (\ref{eq:expansion1}) vanish upon integration. Thus eq. (\ref{eq:expansion1}) simplifies to
\begin{equation}\label{eq:expansion2}
1 + \frac{1}{2!} \Big(\frac{2ap.k}{\Lambda^{2}}\Big)^{2} +  \cdots = \cos{\Big[ \frac{ia\sqrt{p^{2}}k}{\Lambda^{2}} \Big]},\vspace{1mm}
\end{equation}
where we have used the identity $q^{\mu}q^{\nu} = \frac{1}{4}q^{2} g^{\mu\nu}$ to simplify the dot product. Using eq. (\ref{eq:expansion2}), and after performing the Wick rotation and the angular integral, eq. (\ref{eq:NL4pt1}) becomes
\begin{flalign}\label{eq:NL4pt2}
V(p^{2}) & =   -\frac{e^{-\frac{2m^{2}}{\Lambda^{2}}}}{16\pi^{2}} \int_{0}^{1} dx \exp{\Big[\frac{(1-2x+2x^{2})p^{2}}{\Lambda^{2}}}\Big] \nonumber\\
& \times \int_{0}^{\infty} dk\frac{k^{3} e^{-\frac{2k^{2}}{\Lambda^{2}}}}{(k^{2}+\Delta)^{2}}\cos{\Big[ \frac{(1-2x)\sqrt{p^{2}}k}{\Lambda^{2}} \Big]}.
\end{flalign}

The integral is still not doable analytically for $p^{2} \neq 0$, however, in the limit  $p^{2} \ll \Lambda^{2}$, we can approximate it as
\begin{equation}\label{eq:NL4pt3}
V(p^{2}) \simeq   -\frac{e^{-\frac{2m^{2}}{\Lambda^{2}}}}{16\pi^{2}} \int_{0}^{1} dx 
 \int_{0}^{\infty} dk\frac{k^{3} e^{-\frac{2k^{2}}{\Lambda^{2}}}}{(k^{2}+\Delta)^{2}},
\end{equation}
and now this integral is doable exactly for the renormalization condition for the $\phi^{4}$ 2-point function, thus we find for $m^{2} \ll\Lambda^{2}$
\begin{flalign}\label{eq:NL4pt4}
V(p^{2} = 0) & \simeq \frac{1}{32\pi^{2}}e^{-\frac{2m^{2}}{\Lambda^{2}}} \Big[ 1 + \text{Ei}\Big(\frac{-2m^{2}}{\Lambda^{2}} \Big) \Big],\nonumber\\
V(p^{2} = 4m^{2}) &\simeq \frac{-1}{32\pi^{2}}e^{-\frac{2m^{2}}{\Lambda^{2}}} \Big[1 - \text{Ei}\Big(\frac{-2m^{2}}{\Lambda^{2}} \Big)\Big].
\end{flalign}

We have checked that for $\Lambda = 20 m$ (with the on-shell condition assumed), eq. (\ref{eq:NL4pt3}) differs from the exact numerical result by less than $1\%$. Given than the LHC constrains $\Lambda \gtrsim 2.5$ TeV, this approximation is justified.

\end{document}